# ATOMIC PRECISION ADVANCED MANUFACTURING FOR DIGITAL ELECTRONICS

Daniel R. Ward, Scott W. Schmucker, Evan M. Anderson, Ezra Bussmann, Lisa Tracy, Tzu-Ming Lu, Leon N. Maurer, Andrew Baczewski, Deanna M. Campbell, Michael T. Marshall, and Shashank Misra
Sandia National Laboratories, Albuquerque, New Mexico
smisra@sandia.gov

## INTRODUCTION

An exponential increase in the performance of silicon microelectronics and the demand to manufacture in great volumes has created an ecosystem that requires increasingly complex tools to fabricate and characterize the next generation of chips.[1] However, the cost to develop and produce the next generation of these tools has also risen exponentially, to the point where the risk associated with progressing to smaller feature sizes has created pain points throughout the ecosystem.[2] The present challenge includes shrinking the smallest features from nanometers to atoms (10 nm corresponds to 30 silicon atoms). Relaxing the requirement for achieving scalable manufacturing creates the opportunity to evaluate ideas not one or two generations into the future, but at the absolute physical limit of atoms themselves. This article describes recent advances in atomic precision advanced manufacturing (APAM) that open the possibility of exploring opportunities in digital electronics. Doing so will require advancing the complexity of APAM devices and integrating APAM with CMOS.[3-5]

## APAM

The era of atomic manipulation, the ability to place or move single atoms with the spatial precision of single atomic sites in some host material, began in 1989 when Don Eigler used a scanning tunneling microscope (STM) to position 35 Xenon atoms on the surface of nickel to spell out the IBM logo.[6] Similar positioning precision has since been achieved using a broad range of atoms and materials, and different manipulation approaches.[7] However, to date, only one pathway, which we refer to as APAM, has been used to produce functioning atomic precision electronic devices in silicon.[3,4]

APAM places dopants into silicon using surface chemistry, a mechanism not typically used in microfabrication. Examining the process, shown in Fig. 1, reveals some key contrasts with standard processing. The Si (100) surface common to microelectronics has one reactive bond per atom, which can be made unreactive by attaching a hydrogen atom. Starting with a surface fully passivated by hydrogen, which serves as a monatomic resist, an STM can remove hydrogen atoms from silicon atoms site-by-site at high bias, and image the surface at low bias.[8] The hydrogen atom is either present or absent at each site. As a result, the resist cannot be overexposed, in contrast to the resists used in next-generation extreme ultraviolet (EUV) photolithography.[9] The depassivation pattern is transferred into the silicon using surface chemistry. Importantly, APAM has shown the atomically precise placement of dopants and semiconductors, but pathways for oxides and metals have yet to be discovered. Phosphine, a dopant precursor, only attaches to silicon surface sites that have an unsatisfied bond, and not to those sites that remain passivated with hydrogen. Phosphine only produces an electrically active phosphorus donor in windows that are at least six atomic sites in size, which has the effect of error-correcting imperfections in the resist.[10]

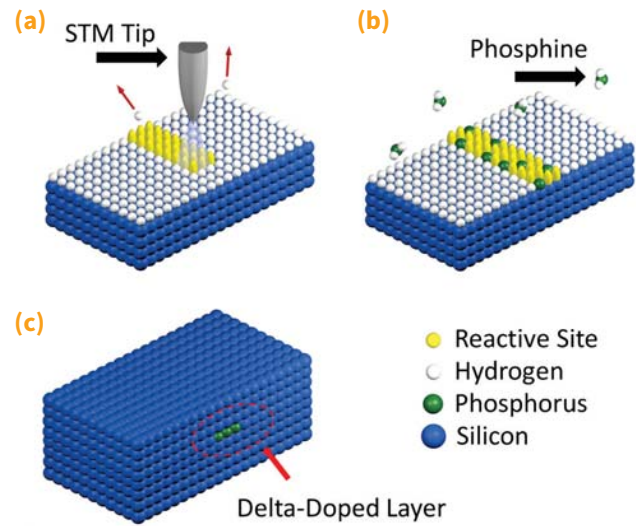

**Fig. 1** APAM process. (a) Schematic showing one hydrogen atom (white) attached to every surface silicon atom (blue). An STM tip is shown removing hydrogen, exposing reactive sites (yellow). (b) Phosphine molecules (containing phosphorus – green) attach selectively only to the reactive sites, and not to hydrogen-terminated sites. (c) The entire two-dimensional phosphorus-based device is encapsulated in silicon to protect it from oxidation and other kinds of damage.

Lastly, capping the surface with a few nm of silicon at moderate temperatures (300°C) satisfies all the bonding to the phosphorus dopant, donating an electron to silicon. The low temperatures enable APAM to achieve higher densities than any other approach— typically $1.7 \times 10^{14}$ cm$^{-2}$ (or over $1 \times 10^{21}$ cm$^{-3}$ in a single atomic plane).[11] Other approaches, such as selective area growth and ion implantation, are limited by the equilibrium concentration of phosphorus in silicon because they require higher temperatures.

Currently, two outstanding challenges inhibit using APAM to develop microelectronics: the need to (1) evolve the complexity of APAM devices relative to the simple devices that have been fabricated so far and (2) integrate with existing CMOS processing (Fig. 2). For both challenges, we will describe the state-of-the-art, the key limitations, and paths forward.

## ADVANCING DEVICE COMPLEXITY

Figure 3 shows a schematic, image, and electrical transport data of an APAM device. This device probes the electronic structure of the two islands between the source and drain leads, each containing only a handful (two to four) of phosphorus donors. The basic characteristics of this device—that it only contains donors, demonstrates basic physics principles, and operates only at cryogenic temperatures (< 4 K)— are common to nearly all state-of-the-art APAM devices.[5] In contrast, digital microelectronics require significantly more complicated device structures, containing both donors and acceptors, with material stacks that permit extensive engineering of the switching performance, all operating at room temperature. Each of these presents a significant challenge.

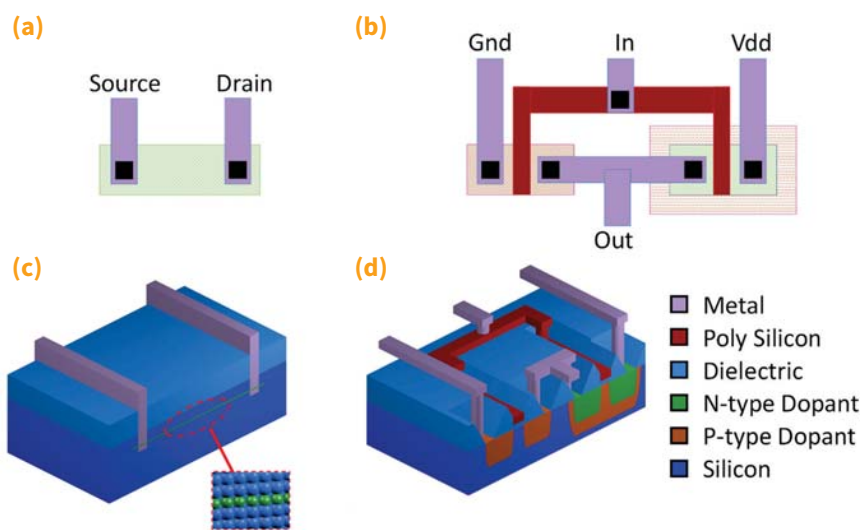

**Fig. 2** Comparison of APAM and CMOS devices. (a,c) Illustration of a typical APAM device showing the buried donor channel (green) and metal source and drain contacts. In-plane donor-based gates are not shown. (b,d) Illustration of a CMOS inverter, showing both an n-type and p-type transistors.

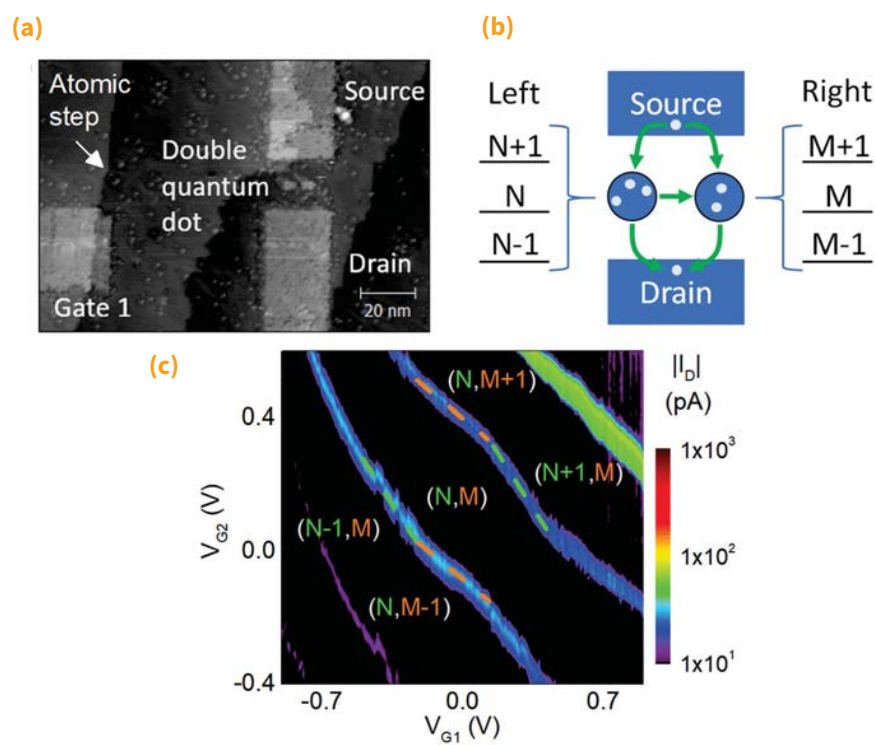

**Fig. 3** APAM-fabricated double quantum dot. (a) STM topographic image showing a pair of ~ 3 nm-wide quantum dots in the channel of a device, and one of two in-plane gates. Bright areas in the image are where hydrogen has been removed. The incongruent, off axis lines running through the image are single-atomic-layer tall step edges. (b) Each of the quantum dots has an integer number of electrons – N for the left quantum dot and M for the right quantum dot. This device conducts in a manner reminiscent of a resonant tunnel diode; namely, when either of the energy levels of the two quantum dots lines up with the energy of the source and drain. (c) False-color image that shows the device passes current when tuning the charge state of the two islands using their respective gates. Green (orange) dashed lines indicate changes in the number of electrons on the left (right) quantum dot. This data was taken at a temperature of 4 K.

Acceptor doping is needed to make complementary transistors (Fig. 4a). While the semiconductor industry employs a number of precursor molecules that are acceptor analogues of phosphine—diborane, boron trifluoride, and trimethyl aluminum, for example—it is unclear whether any can be used with the chemical contrast created by hydrogen-terminated and unterminated silicon. Even if an acceptor can be used, whether they produce a high density of electrically active dopants at temperatures where the hydrogen mask is stable, below 430°C, has not been shown.[10] Early work from IBM-Zurich indicates that diborane satisfies the first of these requirements, with the electrical quality of the material remaining an open question.[12] Another reason for optimism is the development of halogen-based resists, which may provide a different kind of chemical contrast from hydrogen.[13]

Digital electronics rely on the ratio of on and off currents in a transistor to be very large. This requires surface gates with higher gain than the dopant-based structures used historically for this purpose in APAM devices (Fig. 4b). Adding a high-gain surface gate to an APAM device is complicated because phosphorus dopants diffuse above 500°C,[14] destroying the atomic precision of the device. This prevents the use of common dielectrics, like silicon dioxide, which require high growth temperatures. It does leave open low-leakage high-k/ metal gates, although these require complicated multi-layer material stacks (Fig. 4b).[15] In an encouraging recent development, the first fab-compatible surface gate has recently been implemented atop an APAM device.[16]

Another obstacle is realizing device operation at room temperature, which again requires advancing APAM device complexity. At cryogenic temperatures, the density of donors produced by APAM is so high that it provides significant confinement of the electrons, and leakage paths around the dopant layer are frozen out. Room temperature operation requires adding features that confine current to the transistor channel and block unintentional leakage paths (Fig. 4c). For the channel of the transistor itself, confinement can be enhanced by adding acceptors to create a local electric field. Here, the standard preparation method for APAM samples—Joule heating of the silicon—destroys heavily acceptor-doped substrates, so alternative sample preparation techniques are being explored.[17]

## CMOS INTEGRATION

The second outstanding challenge in applying APAM to digital microelectronics is interfacing APAM and CMOS on the same die (Fig. 5).[18,19] While the thermal budget to incorporate activated dopants using APAM is modest, the temperature necessary to prepare the atomically clean surfaces required for APAM has typically been 1200°C. Should APAM come after CMOS fabrication, this temperature would destroy the products of high-temperature steps in a CMOS workflow, the so-called front end of line (FEOL). Similarly, the APAM device only survives up to 500°C, which is insufficient to execute FEOL CMOS manufacturing after APAM. Recently a sample cleaning method operating at 850°C has been demonstrated.[18] This means APAM can fit into a split CMOS fabrication process, wherein high temperature steps such as ion implantation and initial oxidation precede APAM, which is followed by low temperature steps such as oxide deposition and metallization.

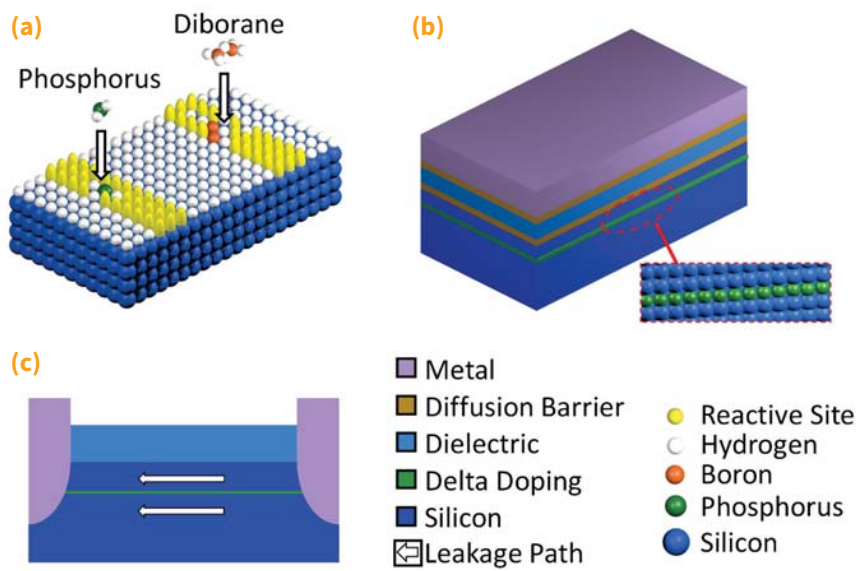

**Fig. 4** **Advancing APAM device complexity. (a) Phosphine is a dopant precursor that contains phosphorus (green), which donates an electron to silicon. Diborane is a dopant precursor that contains boron (orange), which removes (accepts) an electron from silicon. (b) In addition to having metal (violet) and dielectric (light blue), high-k/metal gates also require a diffusion barrier (brown) between the metal-dielectric and dielectric-semiconductor interfaces. (c) This schematic of an APAM device shows current leakage paths (white arrows) through the silicon cap and the substrate that become relevant at room temperature.**

The reduced temperature APAM clean should leave the FEOL CMOS fabrication steps undisturbed, although this has yet to be proven out on real parts. Similarly, back end of line (BEOL) CMOS fabrication steps require low enough temperatures that they should leave the APAM device undisturbed, although that too has yet to be proven out.

An opportunity exists to add APAM features to CMOS circuits even later in fabrication, at the cost of reduced precision (Fig. 6). The temperature required to clean silicon can be reduced to just 100°C by leveraging exotic chemical preparations[17,20] This temperature is less than the operating temperature for most CMOS parts, and opens the door to adding APAM features inside APAM target windows designed into these parts. Wet chemical routes to cleaning silicon do not produce atomically perfect surfaces, but the surface appears clean enough to execute a reduced resolution APAM process. The other steps of the APAM process, from hydrogen termination to dopant incorporation and silicon capping, work with slightly reduced effectiveness at ambient temperatures when starting with atomically perfect surfaces. Patterning the resist becomes a few times less efficient, dopant activation falls from nearly 100% to 75%, and conductivity goes down by a factor of three. These steps have yet to be combined into a single process flow, and the effect of APAM on the functionality of the CMOS host remains unknown, but this approach remains promising.

## PROSPECTUS

Existing data on APAM devices indicates that the work detailed above may well be justified due to the high performance enabled by APAM fabrication. Figure 7 shows that an APAM nanowire can support a current density of up to 2 mA/μm, at 4K, which exceeds modern CMOS, at room temperature. This extremely high current density is likely the result of the high density of carriers, and a sizable confinement potential, both courtesy of the high dopant density. There are reasons to expect much of this performance to be preserved on warming to room temperature. Related quantities such as the carrier density and mobility, as measured by infrared ellipsometry at room temperature, agree with cryogenic Hall measurements to within a factor of two. The current density may be further enhanced by adopting modern approaches, including three-dimensional geometries[21] and engineering channel strain through alloying.[22] Overall, there is every reason to believe that APAM devices can be interfaced directly with CMOS, without any transduction.

Modeling and simulation of APAM devices requires the development of new numerical tools.[23] Most approaches for modeling silicon devices cannot readily accommodate such a high density of dopants with a sharp dopant profile due to limitations of the discretization schemes used for the underlying equations. Moreover, much of the processing relevant for APAM devices

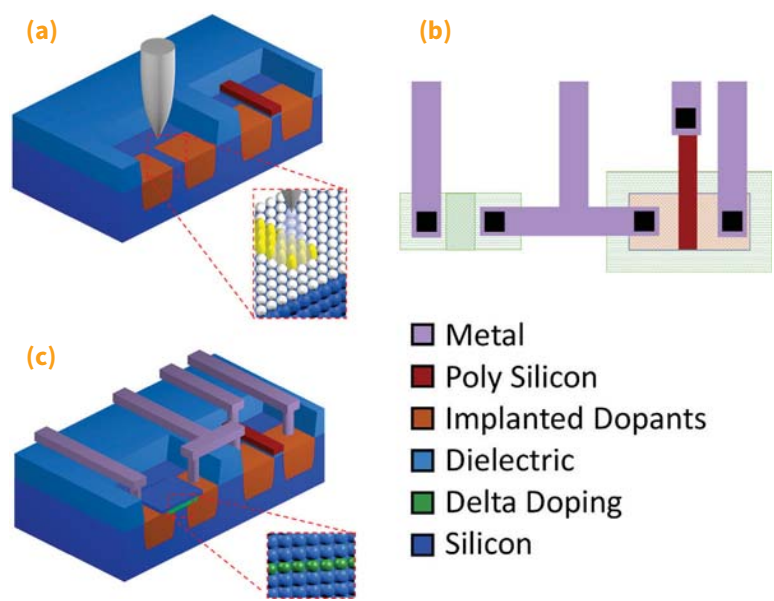

Metal
Poly Silicon
Implanted Dopants
Dielectric
Delta Doping
Silicon

**Fig. 5** Split CMOS integration. (a) CMOS FEOL, which involves high-temperature processing steps, can be done before APAM. These steps include defining a cell (left) to pattern an APAM device on the chip. (b,c) CMOS BEOL, which involves low-temperature processing, can be done after APAM and must not disturb the donor device. These steps include defining contact routing and gates for the APAM cell.

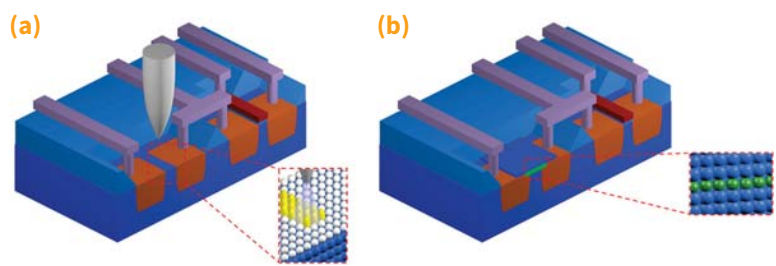

**Fig. 6** Post-CMOS integration. (a) A fully processed CMOS chip has a thermal budget of only around 100-200°C, much lower than the 850°C thermal budget of a CMOS chip only after FEOL (Fig. 5). This is due to the addition of the dielectric (light blue) and metal (purple) layers. After BEOL CMOS processing, a window is opened on the surface of the chip where APAM will be performed. (b) Because of a different order of operations compared to Fig. 5c, the APAM process, including capping silicon, now must not disturb the CMOS chip outside of the window.

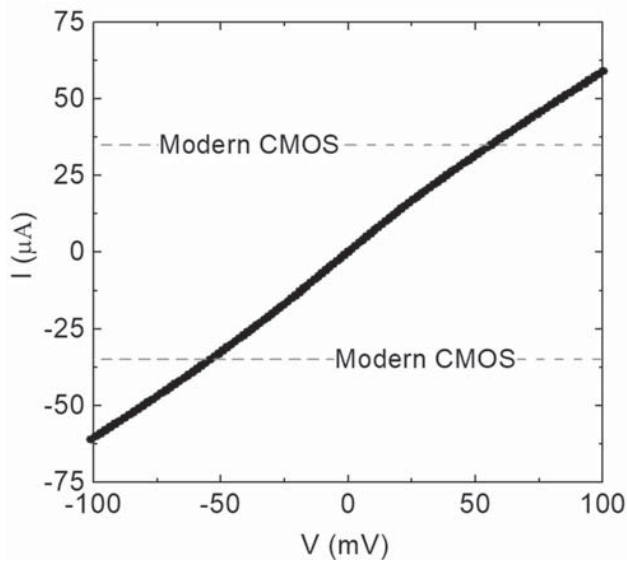

**Fig. 7** APAM current density. The I-V curve for a 35 nm-wide, 105 nm-long nanowire made using APAM, measured at a temperature of 4K. Dash lines indicate typical drive currents for modern CMOS (1 mA/μm) assuming a 35 nm-wide channel.

occurs outside of the useful parameter space of most TCAD (technology computer aided design) process simulators, necessitating extending the applicability of those simulators. Lastly, at the small size-scales, strong confinement, and high current densities that APAM permits access to, quantum effects can no longer be handled semi-classically, even at room temperature. The ability to treat classical and quantum electrical transport on equal footing remains an outstanding problem, and one broadly useful outside of APAM itself.

## CONCLUSION

Advancing the complexity of APAM devices and integrating APAM with CMOS will lead to a powerful tool for advancing microelectronics. At the simplest level, it creates the opportunity to explore the physical limitations of CMOS far into the future. Aside from the ramifications for device function, APAM also provides a platform for studying basic physics such as quantum effects that emerge at small size scales, and statistical number fluctuations that will dominate the behavior of few-atom systems. At a subtler level, it introduces the possibility of changing CMOS processing itself by integrating a higher density of dopants than has been created using any other approach, and at relatively late points of the workflow.

## ACKNOWLEDGMENTS

This work was supported by the Laboratory Directed Research and Development Program at Sandia National Laboratories, and was performed, in part, at the Center for Integrated Nanotechnologies, a U.S. DOE, Office of Basic Energy Sciences user facility.

Sandia National Laboratories is managed and operated by National Technology and Engineering Solutions of Sandia LLC, a wholly owned subsidiary of Honeywell International Inc., for the U.S. Department of Energy under contract DE-NA0003525. The views expressed in the article do not necessarily represent the views of the U.S. DOE or the United States government.

## ABOUT THE AUTHORS

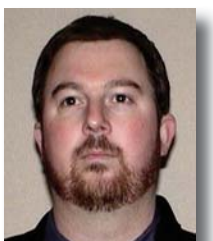

**Dan Ward** is a principal member of technical staff at Sandia National Labs in the Fabrication Technology Development department. He leads fabrication efforts in silicon-based quantum information science and atomically-precise advanced manufacturing. Ward did his Ph.D. in molecular electronics at Rice University and a postdoc in silicon-based quantum computing at the University of Wisconsin-Madison. He is a subject matter expert in electron and ion beam lithography with over ten years of experience in nanolithography. Ward has over 20 publications related to silicon-based quantum computing technology and is highly engaged in the community including fabricating multiqubit devices for academic groups. Recently Ward has been involved in developing more robust fabrication methods for atomically-precise nanostructures.

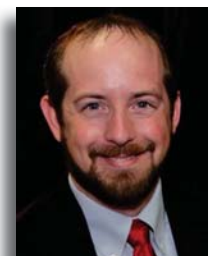

**Scott W. Schmucker** is a senior member of technical staff at Sandia National Laboratories in Albuquerque, New Mexico. He received his master's and doctorate in electrical engineering from the University of Illinois at Urbana-Champaign, and his bachelor's in computer engineering from Case Western Reserve University. He has 15 years of experience in the areas of surface science and atomically precise manufacturing with a focus on scanning tunneling microscopy and hydrogen depassivation lithography. He has co-authored 20 journal articles in these areas and holds two related U.S. patents.

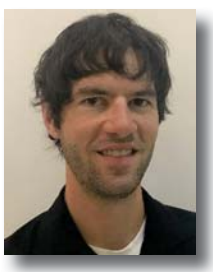

**Evan M. Anderson** earned his bachelor's from Purdue University and his master's and doctorate from the University of Michigan, all in materials science and Engineering. For his graduate research, he conducted molecular beam epitaxy, scanning tunneling microscopy, and density functional theory calculations to investigate the surface structure and alloying of InAsSb. He came to Sandia National Laboratories as a postdoc, and eventually transitioned to fabricating and packaging optoelectronic devices to develop infrared sensors. Most recently, he was hired as a staff member for research and development of Si-based atomic precision electronic devices.

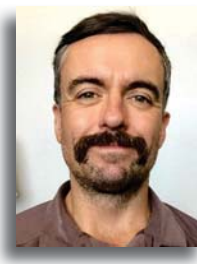

**Ezra Bussmann** earned his bachelor's degree in physics from Beloit College in 2000, and his doctorate from the University of Utah in 2006. He performed postdoctoral research at Sandia National Laboratories and the Centre Interdisciplinaire de Nanoscience de Marseille at Universite Aix-Marseille France. In 2010, he started Sandia's effort in APAM for quantum electronics. His work is in condensed matter physics specializing in structural, electronic, and magnetic properties of thin films and surfaces. The work is captured in 30 journal articles and four patents.

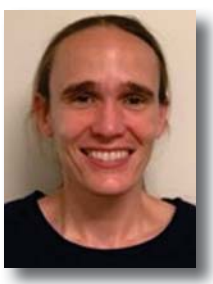

**Lisa Tracy** is currently a principal member of technical staff at Sandia National Labs in the Quantum Phenomena department. She has over 17 years of experience with electrical transport and microwave measurements at dilution refrigerator temperatures and fabrication of semiconductor nanostructures. Tracy obtained her doctorate at the California Institute of Technology, where she used RF and microwave techniques to study low-dimensional systems in semiconductors at cryogenic temperatures. Since 2007, she has been working at Sandia on research in semiconductor nanostructures, focusing on spin qubits.

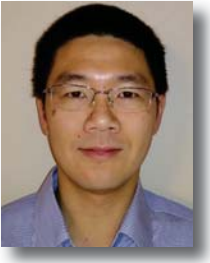

**Tzu-Ming Lu** received his bachelor's from National Taiwan University in 2004 and his doctorate from Princeton University in 2011. After graduate school, he was a postdoctoral researcher at Sandia National Laboratories, New Mexico, where he is currently a senior member of technical staff. His research topics include semiconductor device physics, spin-orbit coupling in solid-state systems, and quantum behavior of nanoscale structures. He is also a Center for Integrated Nanotechnologies (CINT) scientist, supporting user projects on quantum information science and solid-state physics at the nanoscale.

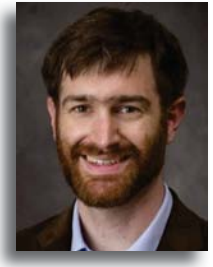

**Leon N. Maurer** received his bachelor's in physics and mathematics from Dartmouth College in 2008 and his doctorate in physics from the University of Wisconsin-Madison in 2016. After graduate school, he was a postdoctoral researcher at Sandia National Laboratories, New Mexico, where he is currently a senior member of technical staff. His current research focuses on modeling a variety of nanoelectronics devices for quantum and classical computing applications. His previous research includes thermal transport in semiconductor nanostructures and superconducting quantum computing.

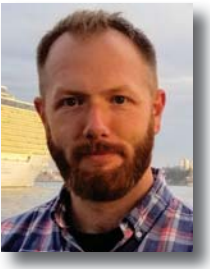

**Andrew Baczewski** is a principal member of technical staff in the Quantum Computer Science group at Sandia National Laboratories. In 2007, he received a bachelor's degree in electrical engineering from Michigan State University. In 2013, he received an interdisciplinary doctorate in physics and electrical engineering from Michigan State University where he was a National Science Foundation Graduate Research Fellow. He then started as a postdoc at Sandia, where he has been ever since. His research interests span computational physics, electronic structure theory, and quantum information science.

**DeAnna Campbell** is a member of the technical staff at Sandia National Laboratories in the Nano and Microsensors Department and received her chemical engineering degree from the University of New Mexico. She has 10 years of experience in sensor technologies including field deployable biosensors utilizing cell-directed assembly of lipid-silica nanostructures, multiplexed microneedle-based biosensor arrays, and shear horizontal surface acoustic wave sensors for small molecule detection. For the past five years, Campbell has worked in the fields of silicon-based quantum information science and atomic precision manufacturing.

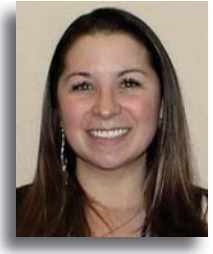

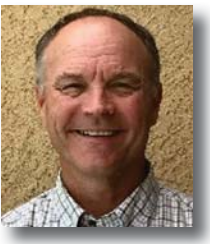

**Michael T. Marshall** received a bachelor's in mechanical engineering and a master's in material science, both from the University of Illinois at Urbana-Champaign (UIUC). His career began at the United States Gypsum Corp.'s product research facility. He then returned to UIUC as a research engineer in the Center for Microanalysis of Materials specializing in TEM, FIB, and microfabrication. After 21 years, he moved to Sandia National Laboratories where he currently is involved with prototyping the next generation electron accelerator system.

**Shashank Misra** earned a doctorate in physics from the University of Illinois – Urbana, Champaign in 2005, and since 2013, has been a member of the research staff at Sandia National Laboratories. His research interests have revolved around developing instruments, techniques, and devices that provide new access to exotic phases in quantum materials, quantum phase transitions, and quantum effects in semiconductors. More recently, his interests have turned to using chemical vapor deposition and STM-based lithography to fabricate atomically-precise dopant devices in semiconductors. He leads the Far-Reaching Applications, Implications, and Realization of Digital Electronics at the Atomic Limit (FAIR DEAL) program at Sandia.

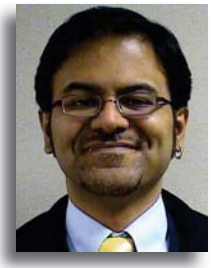